\documentclass[12pt]{article}
\pdfoutput=1

\usepackage{epsfig,latexsym,amsfonts,amsmath,amsthm,amssymb,amsbsy,multirow,slashed,wasysym,textcomp,subfigure,hyperref,wrapfig}
\usepackage[font={footnotesize,sl},bf]{caption}

\usepackage{xcolor}

\def\be{\begin{equation}}
\def\ee{\end{equation}}
\def\bea{\begin{eqnarray}}
\def\eea{\end{eqnarray}}

\def\nn{\nonumber}

\numberwithin{equation}{section}

\newcommand{\caltA}{{\cal{\widetilde {A}}}}
\newcommand{\tZ}{\widetilde {Z}}
\newcommand{\tk}{\widetilde {k}}
\newcommand{\ta}{\widetilde {a}}
\newcommand{\tA}{\widetilde {A}}
\newcommand{\tP}{\widetilde {P}}

\newcommand{\tW}{\widetilde {W}}

\newcommand{\tV}{\widetilde {V}}
\newcommand{\g}{\gamma}
\newcommand{\tmu}{\widetilde {\mu}}
\newcommand{\te}{\widetilde {e}}

\def\g{\gamma}

\def\qKK{q_{\text{\tiny{KK}}}}

\def\lUV{ l_{\text{\tiny{UV}}}}

\numberwithin{equation}{section}

\begin{document}
 
\begin{titlepage}

\begin{flushright}
IPhT-T15/184\\
CPHT-RR043.1015\\
ULB-TH/15-20\\
\end{flushright}
\bigskip
\bigskip

\centering{\Large \bf $AdS_3$: the NHEK generation}

\bigskip\bigskip\bigskip

 \centerline{{\bf Iosif Bena,}\footnote{\tt iosif.bena@cea.fr}${}^\dagger$ 
 {\bf Lucien Heurtier,}\footnote{\tt  lucien.heurtier@cpht.polytechnique.fr}${}^\diamondsuit{}^\clubsuit$ and {\bf Andrea Puhm}\footnote{\tt puhma@physics.ucsb.edu}${}^\spadesuit$}
 
 \bigskip \bigskip
\centerline{\em ${}^\dagger$Institut de Physique Th\'eorique, CEA Saclay,}
\centerline{\em 91191 Gif sur Yvette, France}
\bigskip
\centerline{\em ${}^\diamondsuit$Centre de Physique Th\'eorique, \'Ecole Polytechnique, CNRS, }
\centerline{\em Universit\'e Paris-Saclay, F-91128 Palaiseau, France}
\bigskip
\centerline{\em ${}^{\clubsuit}$Service de Physique Th\'eorique, Universit\'e Libre de Bruxelles,}
\centerline{\em Boulevard du Triomphe, CP225, 1050 Brussels, Belgium}
\bigskip
\centerline{\em ${}^\spadesuit$Department of Physics, University of California,}
\centerline{\em Santa Barbara, CA 93106 USA}
\bigskip

\begin{abstract}

It was argued in~\cite{Bena:2012wc} that the five-dimensional near-horizon extremal Kerr (NHEK)  geometry can be embedded in String Theory as the infrared region of an infinite family of non-supersymmetric geometries that have D1, D5, momentum and KK monopole charges. We show that there exists a method to embed these geometries into asymptotically-$AdS_3 \times S^3/{\mathbb{Z}}_N$  solutions, and hence to obtain infinite families of flows whose infrared is NHEK. This indicates that the CFT dual to the NHEK geometry is the IR fixed point of a Renormalization Group flow from a known local UV CFT and opens the door to its explicit construction.

\end{abstract}

\end{titlepage}

\setcounter{tocdepth}{2}
\tableofcontents

\section{Introduction}

The Kerr-CFT conjecture~\cite{Guica:2008mu} relates the near-horizon geometry of an extremal Kerr black hole \cite{Bardeen:1999px} to a putative 1+1 dimensional conformal field theory whose central charges are given by the angular momenta of this black hole. This connection allows one to count microscopically the entropy of extremal Kerr black holes, and its discovery is an important step in extending the powerful machinery of string theory to analyze non-supersymmetric black holes. 

Unfortunately, besides the central charge, very little is known about the CFT dual to the Near-Horizon Extremal Kerr (NHEK) geometry. In particular, it is not known whether this CFT can be realized as the infrared fixed point of a Renormalization Group (RG) flow from a known UV theory, or as a low-energy theory on a system of strings and branes. In order to achieve such a construction it is important to embed the NHEK geometry in String Theory, and to look for a system of branes that, when gravity is turned on, gives rise to a NHEK geometry. The first attempt at such an embedding was made by the authors of~\cite{Guica:2010ej, Song:2011ii}, who constructed a solution with D1, D5 and Taub-NUT charges, that has a five-dimensional NHEK geometry in the infrared. However, this solution has fixed moduli, and hence does not allow one to search for a microscopic theory that flows in the infrared to the CFT dual to NHEK.

In \cite{Bena:2012wc}, it was shown that one can embed the five-dimensional NHEK geometry in a very large family of supergravity solutions, parameterized by several continuous parameters, and that moreover one can obtain multicenter solutions where the geometry near one of the centers is NHEK. All these solutions belong to a class of extremal non-supersymmetric solutions that can be obtained by performing a duality sequence  known as generalized spectral flow \cite{Bena:2008wt} on the well-known almost-BPS solutions \cite{Goldstein:2008fq,Bena:2009ev}. The existence of this very large family of solutions that have a NHEK region in the infrared raises the hope that one may be able to find a flow from a UV that is $AdS$ to the NHEK geometry in the infrared. This would imply that the UV CFT is a ``nice'' local CFT, with well-understood 
operators, etc. One could then go ahead and investigate this UV CFT and find which operators trigger the RG flow to the NHEK solution. 

This paper has two main goals. The most important one is to develop two systematic procedures to embed the infinite families of multicenter solutions with NHEK regions~\cite{Bena:2012wc} into asymptotically-$AdS_3$ solutions. The first procedure is to use the explicit form of the solutions~\cite{Bena:2012wc}   and to investigate various limits of the parameters that control the UV of these solutions in order to produce an $AdS$ factor in the metric. The second is to make clever use of the fact that the only difference between the asymptotics of BPS and almost-BPS solutions is the sign of one of the electric fields~\cite{Goldstein:2008fq}, of the fact that BPS solutions transform under generalized spectral flows into other BPS solutions~\cite{Bena:2008wt}, and of the fact that BPS solutions develop an $AdS_3$ UV region when certain of their moduli are put to zero, in order to perform a systematic search for solutions with a NHEK infrared and an $AdS_3$ UV. 

At first glance, both these procedures should be automatically successful. The asymptotics of the solutions is controlled by 17 parameters, and if one sets to zero the constant, $1\over \rho$ and $\cos {\theta} \over \rho^2$ terms in $g_{\rho \rho}$ and certain divergent components of the fields one should obtain a 10-parameter family of solutions that have the leading radial component of the metric of the form $d \rho^2 \over \rho^2$, which is the hallmark of an asymptotically-$AdS$ solution. However, things are not so simple. Most of the solutions obtained in this class have closed timelike curves, and if one tries to naively impose all the conditions that eliminate the closed timelike curves none of the asymptotically-$AdS$ solution seem to survive. Similarly, if one tries to obtain these solutions by relating BPS and almost-BPS solutions (as we will explain in detail in Appendix~\ref{app:BPStrick}) at the end one has to solve 7 equations for 17 variables, which however appear at first glance to have no 
solution. Thus, despite the presence of a large number of available constants, neither of the two hunting methods we use seems to be very willing to yield asymptotically-$AdS$ solutions.

Fortunately, a careful analysis reveals that things are not so bleak. Indeed, we find that among the many ways to solve the constraints associated to the absence of closed timelike curves, there is one that produces nontrivial solutions that are asymptotically $AdS_3 \times S^3/{\mathbb{Z}}_N$. Furthermore, the second method produces exactly the same solution, which we take as a remarkable confirmation that we have really identified {\it the} way to construct solutions that have an $AdS_3$ factor in the UV and NHEK in the IR. The existence of such a class of flows has several important implications. 

The first is for the debate whether the theory dual to the NHEK geometry can be described as the infrared limit of a local CFT$_2$ or only of a non-local one. Since the NHEK geometry can be obtained by a certain identification of an (uncompactified) warped $AdS_3$ geometry, it was argued in~\cite{ElShowk:2011cm} that the theory dual to NHEK is the DLCQ of a nonlocal theory which is dual to warped $AdS_3$ (oftentimes known as a ``dipole'' quantum field theory). Another proposal is that the dual to NHEK is given by the identification of a more exotic type of conformal field theory, called warped-CFT (wCFT)~\cite{Hofman:2011zj,Azeyanagi:2012zd,Detournay:2012pc}, which would be a local theory  \cite{Hofman:2014loa}. Our construction shows that the theory dual to the NHEK geometry could be equally well obtained as the IR fixed point of many RG flows of ``vanilla CFT's'' and hence it can be understood without resorting to wCFT's or dipole theories. 
 
Second, this class of solutions opens the door to identifying the embedding of the Virasoro symmetry of the CFT dual to the NHEK geometry in the Virasoro symmetry of the UV CFT, by finding a diffeomorphism that interpolates between Brown-Henneaux diffeomorphisms \cite{Brown:1986nw} of the UV and the IR, similar to the construction of~\cite{Compere:2015bca}. Third, this family of flows should allow one to understand which deformation of the D1-D5 CFT one needs to turn on to flow to a  NHEK infrared\footnote{And conversely what is the deformation of the NHEK geometry that allows it to {\it UV-flow}  to a nice local CFT in the UV~\cite{ElShowk:2011cm}.}, and in particular whether this deformation is similar to the one that triggers the RG flow of the asymptotically $AdS_3 \times S^3$ solution to the near-horizon $AdS_3$ of a BPS black ring \cite{Bena:2004tk}, or whether it is rather a deformation of the Lagrangian. Thus, the existence of this family of flows opens a new route for determining what the CFT dual to 
5D NHEK is. 
 
In addition to embedding the NHEK-containing solutions of \cite{Bena:2012wc} in $AdS_3$,  we also construct the full form of their R-R three-form field strength. These solutions can be obtained by dualizing the twice-spectrally-flowed almost BPS solutions that were constructed in the M2-M2-M2 duality frame in \cite{Dall'Agata:2010dy} to the D1-D5-P duality frame. In \cite{Bena:2012wc} this duality transformation was performed for the metric and the dilaton, which was enough to ascertain the existence of these solutions and to perform some basic regularity checks. However,  to perform all the regularity checks and to be able to understand all the properties of these solutions one must also construct this three-form explicitly.  As we will see in Appendix~\ref{app:F3}, even if this construction involves several duality transformations that act rather nontrivially on the R-R fields, the final implicit form of the expression that give three-form field strength is 
quite simple. However, its explicit form is much more complicated than for BPS and almost-BPS solutions, even after making several simplifying assumptions (equation \eqref{eq:monster}). 

Besides its importance for the programme of embedding the NHEK geometry in String Theory, the calculation of the R-R three-form also fills an important gap in our knowledge of almost-BPS solutions and generalized spectral flows thereof. Indeed, the full solution that comes from applying three generalized spectral flows on an almost-BPS solution has so far only been constructed in the M2-M2-M2 duality frame \cite{Dall'Agata:2010dy}. Writing some of these solutions in the D1-D5-P duality frame allows one to embed these solutions into six-dimensional ungauged supergravity and explore whether these solutions belong to a larger class of wiggly solutions, as it happens when supersymmetry is preserved~\cite{Niehoff:2013kia,Bena:2015bea}.
\\

This paper is organized as follows. In~\S~\ref{sec:NHEKembedding} we review the family of supergravity solutions that contain NHEK regions in the infrared and that can be obtained by a sequence of generalized spectral flow transformations from almost-BPS solutions with D1, D5, momentum and KK monopole charges. We then find the explicit R-R fields for these solutions whose derivation is given in Appendix~\ref{app:F3}. In~\S~\ref{sec:UV} and Appendix~\ref{app:BPStrick} we develop two different systematic procedures to search for solutions with an $AdS$ UV, and identify a sub-class of the large family of supergravity backgrounds with a NHEK infrared constructed in~\cite{Bena:2012wc} that have an  $AdS_3 \times S^3/\mathbb{Z}_N$ asymptotic region.

\section{Infinite families of NHEK embeddings in String Theory} \label{sec:NHEKembedding}

In~\cite{Bena:2012wc} it was shown that an infinite family of IIB supergravity solutions with a NHEK infrared can be obtained by performing generalized spectral flow transformations \cite{Bena:2008wt,Dall'Agata:2010dy} on a class of non-supersymmetric, ``almost-BPS'', multicenter supergravity solutions \cite{Goldstein:2008fq,Bena:2009ev} whose charges correspond to D1 and D5 branes, momentum and KK monopoles. However, in~\cite{Bena:2012wc} only the metric and dilaton have been constructed. While this was enough to ascertain the existence of such solutions, in order to perform all regularity checks, calculate their asymptotic charges or use holography to read off the features of the UV CFT, the explicit expressions for the R-R fields are needed. It is the purpose of this section to complete the construction of the supergravity solutions that contain a NHEK infrared by explicitly computing these R-R fields. 

\subsection{Almost-BPS D1-D5-P-KK solutions}\label{subsec:almostD1D5pKK}

The metric of the extremal BPS and almost-BPS D1-D5-P-KK solutions is~\cite{Elvang:2004ds,Bena:2004tk,Bena:2008dw}\footnote{We work in units where $\alpha'=g_s=1$ and use the conventions of \cite{Dall'Agata:2010dy}. This solution can be derived from a sequence of duality transformation acting on an M2-M2-M2 solution where the three M2 branes wrap different two-tori $T^2_I$. The notation using subscripts $I=1,2,3$ is a vestige of this M-theory solution. Note that the relative sign in the metric~\eqref{eq:metricMtoIIB} between $dz$ and $A_3$ differs from the one in~\cite{Bena:2008dw}. We explain the reason for this in appendix~\ref{app:F3} where we explicitly carry out the sequence of duality transformations on the spectrally flowed M2-M2-M2 solution to get the spectrally flowed D1-D5-P solution reviewed in~\S~\ref{subsec:completion}.}
\begin{equation}
  d  s^2= - \frac{1}{ {Z}_3\sqrt{ {Z}_1 {Z}_2}}( d  t +  {k})^2 + \sqrt{ {Z}_1 {Z}_2}  d s_4^2 + \frac{{Z}_3}{\sqrt{ {Z}_1 {Z}_2}}\left(  {A}_3 -dz\right)^2+ \sqrt{\frac{ {Z}_1}{ {Z}_2}} ds_{T^4}^2\,,\label{eq:metricMtoIIB}
\end{equation}
where $z=-R_y y -t$ with $R_y$ the radius of the $y$ circle wrapped by the D1 and D5 branes and carrying momentum and $y$ is periodically identified with $2\pi$. 
The dilaton is given by $e^{2\Phi}= \frac{Z_1}{Z_2}$ 
and the  field $B^{(2)}=0$. 
The R-R three-form flux is given by
\begin{equation}\label{eq:fluxMtoIIB}
F^{(3)}= dA_1 \wedge (A_3 -dz) -\left(\frac{ {Z}^5_2}{{Z}^2_3 {Z}^3_1}\right)^{1/4}\star_5 d A_2 \,.
\end{equation}
The five-dimensional Hodge star is with respect to the time-fibration over the four-dimensional base space
\begin{equation}
ds_4^2=V^{-1} (d\psi+A)^2+V(d\rho^2 +\rho^2 (d\theta^2+\sin^2\theta d\phi^2)) \qquad \text{with} \quad  \star_3 dA=\pm dV\,,\label{eq:TN}
\end{equation}
where the three-dimensional Hodge star is with respect to the three-dimensional flat space. The one-form $A$ is a Kaluza-Klein gauge field and the function $V$ is the Taub-NUT potential:
\begin{equation}
 V=v_0 + \frac{\qKK}{\rho}\,.
\end{equation}
The sign $\pm$ in \eqref{eq:TN} specifies the orientation of the Taub-NUT base and distinguishes between BPS and almost-BPS solutions \cite{Goldstein:2008fq}. We will consider almost-BPS solutions corresponding to the minus sign in~\eqref{eq:TN} for which  $A=-\qKK\cos\theta d\phi$.
The one-forms $A_I$ consist of an ``electric'' and a ``magnetic'' part:
\begin{equation}
A_I=- \frac{dt+k}{Z_I}+a_I\,,
\end{equation}
with the warp factors $Z_I$ with $I=1,2,3$ encoding respectively the asymptotic electric D1, D5 and momentum charges and the vector potentials $a_I$ encoding the local magnetic dipole charges.
In the base space \eqref{eq:TN} the magnetic one-form potentials $a_I$ and the angular momentum one-form $k$ can be decomposed as
\begin{equation}
 a_I = K_I (d\psi +A) + w_I\,, \quad\quad k=\mu (d\psi+A) + \omega\,,
\end{equation}
where $K_I$ and $\mu$ are functions and $w_I$ and $\omega$ are one-forms in the three-dimensional flat space.
Solution specified by $Z_I, a_I, k$ are obtained by solving the almost-BPS equations
\begin{eqnarray}
 d\star_3 d Z_I &=& \frac{C_{IJK}}{2} V d\star_3 d(K_J K_K)\,,\label{eq:almostBPS0}\\
 \star_3 dw_I &=& K_I dV - V dK_I\,,\label{eq:almostBPS1}\\
 \star_3 d\omega &=& d(\mu V) - V Z_I dK_I\,,\label{eq:almostBPS2}
\end{eqnarray}
where $C_{IJK}=|\epsilon_{IJK}|$.
Acting with $d\star_3$ on \eqref{eq:almostBPS2} yields an equation for $\mu$:
\begin{equation}
 d\star_3 d(V \mu) = d(V Z_I) \star_3 dK_I\,.\label{eq:almostBPSmu}
\end{equation}
The solutions $Z_I$ and $\mu$ to \eqref{eq:almostBPS0} and \eqref{eq:almostBPSmu} contain harmonic functions which we will denote by $L_I$ and $M$. An almost-BPS solution is then determined by the functions $(V,\{K_I\},\{L_I\},M)$.
In anticipation of~\S~\ref{subsec:completion} we also define~\cite{Dall'Agata:2010dy} the one-forms $v_I$ and $\nu$ obtained by solving\footnote{Note that we renamed the one-form $v_0$ of~\cite{Dall'Agata:2010dy} to $\nu$ since we already use $v_0$ as the constant in the Taub-NUT potential.}
\begin{eqnarray}
  \star_3 dv_I &=& dZ_I - \frac{C_{IJK}}{2} \left[ V d(K_J K_K) - K_J K_K dV\right]\,,\label{eq:almostBPS3}\\
 \star_3 d\nu &=& Z_I dK_I -K_I dZ_I + V d (K_1 K_2 K_3) - K_1 K_2 K_3 dV\,.\label{eq:almostBPS4}
\end{eqnarray}

\subsection{Non-BPS D1-D5-P-KK solutions with a NHEK infrared}\label{subsec:completion}

Applying a sequence of supergravity transformations known as generalized spectral flows~\cite{Bena:2008wt} to the solution~\eqref{eq:metricMtoIIB}-\eqref{eq:fluxMtoIIB} yields a large class of solutions that contains the NHEK geometry as an infrared limit. We refer to~\cite{Dall'Agata:2010dy} for details about the generalized spectral flow transformations and summarize here the solution for the NS-NS fields~\cite{Bena:2012wc}. We then go on to compute the R-R fields of the spectrally flowed solution.


The spectrally flowed extremal D1-D5-P-KK metric is~\cite{Dall'Agata:2010dy}:
\begin{equation}\label{eq:widetildeds}
  d \widetilde s^2= - \frac{1}{\widetilde {Z}_3\sqrt{\widetilde {Z}_1\widetilde {Z}_2}}( d  t + \widetilde {k})^2 + \sqrt{\widetilde {Z}_1\widetilde {Z}_2}  d\widetilde s_4^2 + \frac{\widetilde {Z}_3}{\sqrt{\widetilde {Z}_1\widetilde {Z}_2}}\left( \widetilde {A}_3 -dz\right)^2+ \sqrt{\frac{\widetilde {Z}_1}{\widetilde {Z}_2}}ds_{T^4}^2\,.
\end{equation}
The dilaton is \mbox{$e^{2\Phi}=\frac{\widetilde {Z}_1}{\widetilde {Z}_2}$} and the NS-NS field $B^{(2)}=0$. 
The four-dimensional base is
\begin{equation}\label{eq:widetildeds4}
 d\widetilde s_4^2 = \widetilde{V}^{-1}(d\psi + \widetilde{A})^2 + \widetilde{V} (d\rho^2 +\rho^2 (d\theta^2+\sin^2\theta d\phi^2))\,.
 \end{equation}
The one-form gauge potentials are given by
\begin{equation}\label{eq:AI}
 \tA_I=-\frac{dt+\tk}{\tW_I}+\ta_I\,,
\end{equation}
and the magnetic one-form potentials $\widetilde a_I$ and the angular momentum one-form $\tk$ can be decomposed as
 \begin{equation}\label{eq:tildeak}
\widetilde a_I=\tP_I(d\psi+\widetilde A)+\widetilde w_I\,, \quad \quad \widetilde{k} = \widetilde{\mu}(d\psi+\widetilde{A}) + \widetilde\omega \,.
\end{equation}
with $\widetilde \omega = \omega$.
The one-forms $\widetilde w_I$ and $\tA$ in the three-dimensional flat space are given by
\begin{equation}\label{eq:tildeoneforms}
\widetilde{A} = A - \gamma_I \, w_I - \frac{C_{IJK}}{2} \gamma_J \gamma_K \, v_I + \gamma^3 \, \nu\,,\quad
\widetilde{w}_I = w_I + C_{IJK} \gamma_J \, v_K - \frac{C_{IJK}}{2} \gamma_J \gamma_K \, \nu \,,
\end{equation}
and functions $\tZ_I, \tW_I, \tP_I,\tmu$ and $\tV$ are given by
\begin{eqnarray}\label{eq:tildefunctions}
 \widetilde{Z}_I &=& \frac{N_I}{\widetilde{V}} \,, \nonumber\\
  \widetilde{W}_I &=& \frac{N_I}{T^3 V + \frac{C_{IJK}}{2}\gamma_J \gamma_K \, T_I Z_I - C_{IJK} \gamma_I \gamma_J \, T_K Z_K} \,, \nonumber\\
  \widetilde{P}_I &=&  \frac{1}{N_I} \left[ V Z_I T_I K_I + \frac{C_{IJK}}{2} \gamma_I \, Z_J Z_K -(2 T_1 -1 ) V \mu \right]+ \frac{\widetilde{\mu}}{\widetilde{W}_I}\,, \\
 \widetilde{\mu} &=&  \frac{1}{\widetilde{V}^2}\left( -\gamma^3 \, Z^3 + {C_{IJK} \over 2} \gamma_J \gamma_K \, Z_I T_I V \mu - {C_{IJK} \over 2} \gamma_I \, V T_J Z_J T_K Z_K + T^3 V^2 \mu \right) , \nonumber\\
 \widetilde{V} &=& \left[ {C_{IJK} \over 2} \gamma_J^2 \gamma_K^2 T_I^2 Z_I^2 - C_{IJK} \gamma_I^2 \gamma_J \gamma_K T_J Z_J T_K Z_K\right. \nonumber\\ 
&& \quad \quad \quad\left. - T^3 V (C_{IJK} \gamma_J \gamma_K T_I Z_I) + T^6 V^2 + 8 \gamma^3 T^3 V \mu \right]^{1/2}  \,, \nonumber
\end{eqnarray}
where $T=(T_1 T_2 T_3)^{1/3}$ and $\gamma=(\gamma_1 \gamma_2 \gamma_3)^{1/3}$ and we defined 
\begin{equation}
 T_I=1+ \g_I K_I\,,\quad \quad N_I= \frac{C_{IJK}}{2} \g_I^2 Z_J Z_K + VT_I^2 Z_I-2 \g_I V T_I \mu\,.
\end{equation}

In order to complete the solution of~\cite{Bena:2012wc} we need to construct the R-R three-form flux of the spectrally-flowed solution, by performing a series of duality transformations that act rather non-trivially on the R-R fields and  are summarized Appendix~\ref{app:F3}. The implicit expression for this flux has a similar form as that of almost-BPS solutions:
\begin{equation}\label{eq:widetildeF3}
\widetilde F^{(3)}=   d  \widetilde A_1 \wedge (\widetilde A_3 -dz) -\left(\frac{\widetilde {Z}^5_2}{\widetilde {Z}^2_3\widetilde {Z}^3_1}\right)^{1/4}\star_5 d\widetilde A_2 \,,
\end{equation}
where the five-dimensional Hodge star is with respect to the time-fibration over the four-dimensional base space~\eqref{eq:widetildeds4}.

\subsection{Towards the full explicit form of the R-R fields}

Despite its apparent simplicity, the explicit expression of  $\widetilde F^{(3)}$ in terms of the harmonic functions determining the solutions is a very complicated nested expression. In particular the action of the five-dimensional Hodge star  involves repeated use of several of the tilded forms and functions and the application of the almost-BPS equations~\eqref{eq:almostBPS0}~-~\eqref{eq:almostBPS4}. 

The purpose of this subsection is to give the complete explicit form of this field for a certain sub-class of solutions.  We will restrict ourselves to solutions with $\mu=0$ (yet keeping $\omega\neq0$), that only have constant terms in the $K_I$ harmonic functions (corresponding to Wilson lines along the Taub-NUT direction in five dimensions and to axion vev's in four dimensions) but no poles. The class of solutions whose explicit three-form field we find does not include the asymptotically-$AdS_3$ solutions that are the main focus of this paper. However, we hope that the (rather complicated) expression we find will be an important stepping stone for finding the three-form of that more complicated class of solutions.

To facilitate the calculation we note that while the first two generalized spectral flows with parameters $\gamma_1,\gamma_2$ act non-trivially on the solution the third spectral flow with parameter $\gamma_3$ corresponds to a coordinate transformation. Hence, without loss of generality we can set $\gamma_3=0$ (which implies that $T_3=1$).
To obtain an explicit expression for the three-form flux from~\eqref{eq:widetildeF3} we express all five-dimensional Hodge stars in terms of three-dimensional ones and make successive use of the almost-BPS equations~\eqref{eq:almostBPS0}~-~\eqref{eq:almostBPS4}. We refer to Appendix~\ref{app:F3} for the details.

The explicit expression for the three-form flux is:
\begin{align}\label{eq:monster}
&\widetilde F^{(3)} = 
 \left[(dt+\omega)\wedge d\left(\frac{T_1 T_2 V - \g_1 \g_2 Z_3}{N_1}\right) - (d\psi+\tA) \wedge d\Big(\frac{T_1 K_1 Z_1 V+\g_1 Z_2 Z_3}{N_1}\Big) \right]  \wedge (\widetilde w_3-dz)  \nonumber\\
& +  {(dt+\omega) \wedge (d\psi+\tA)} \wedge \left[\frac{T_1 T_2 V+\g_1 \g_2 Z_3}{N_3}{d\left(\frac{T_1 K_1 Z_1 V+\g_1 Z_2 Z_3}{N_1}\right)} - K_3 d\left(\frac{|T_1 T_2 V - \g_1 \g_2 Z_3|}{N_1}\right)\right] \nonumber\\
& + \left[ {(\widetilde w_3-dz)}-\frac{T_1 T_2 V+\g_1 \g_2 Z_3}{N_3} (dt+\omega) +K_3  {(d\psi+\tA)}\right] \wedge \left[\frac{T_1 K_1 Z_1 V+\g_1 Z_2 Z_3}{N_1}  {d\tA}+ {d\widetilde w_1}\right]\nonumber\\
& + {(dt+\tk)}\wedge \Big[ \frac{\g_2 T_2}{N_3}  (V  {\star_3 dZ_3}-Z_3  {\star_3 dV}) +\frac{T_1 T_2 V - \g_1 \g_2 Z_3}{N_1}(\g_1 T_2  {\star_3 dZ_2} + \g_2 T_1  {\star_3 dZ_1})\Big]  \nonumber\\
& +(d\psi+\tA)\wedge \Big[\frac{\g_1 \g_2 T_2^2 Z_2+\g_2^2 T_1 T_2 Z_1}{(T_1 T_2 V - \g_1 \g_2 Z_3)^2}(V  {\star_3 dZ_3}-Z_3  {\star_3 dV}) +\frac{\g_2^2 Z_3  {\star_3 dZ_1}+T_2^2 V  {\star_3 dZ_2}}{T_1 T_2 V - \g_1 \g_2 Z_3}\Big]\nonumber\\
& +(dt+\omega)\wedge(d\psi+\tA) \wedge \frac{\g_2 T_1 Z_1 Z_3  {dV}- \g_1 T_2 Z_2 V  {dZ_3}}{N_1 N_3}\,,
\end{align}
where
\begin{align}
\begin{split}
& \qquad\quad \qquad \widetilde w_1=w_1+\g_2 v_3\,, \quad \widetilde w_3=w_3+\g_1 v_2 +\g_2 v_1 - \g_1\g_2\nu\,, \\
&\tk=-\frac{V Z_3(\g_1T_2Z_2+\g_2 T_1Z_1)}{\tV^2}(d\psi + \tA)\,,\quad \tA=A-\g_1 w_1 - \g_2 w_2 - \g_1\g_2 v_3\,,
\end{split}
\end{align}
and
\begin{eqnarray}
  N_1=\g_1^2 Z_2 Z_3 + VT_1^2 Z_1\,,\quad N_2=\g_2^2 Z_1 Z_3+VT_2^2Z_2\,,\quad N_3 = VZ_3\,.
\end{eqnarray}
It is a straightforward although tedious exercise to check that this field strength is closed.

\section{An $AdS_3$ throat with a NHEK}\label{sec:UV}

We now explore the asymptotics of the supergravity solutions of~\S~\ref{sec:NHEKembedding} and, in particular, whether it is possible to construct a geometry with an $AdS_3$ ultraviolet and a NHEK infrared. We will pursue two strategies. The first is to investigate various limits of the parameters controlling the UV of these solutions in order to produce an $AdS$ factor in the metric. The asymptotics of the solutions is controlled by 17 parameters. Setting to zero the constant, $\frac{1}{\rho}$, $\frac{\cos \theta}{\rho}$ and $\frac{\cos\theta}{\rho^2}$ terms in $g_{\rho \rho}$ and certain divergent components of the fields should yield a 10-parameter family of solutions whose leading radial component of the metric is of the form $\frac{d\rho^2}{\rho^2}$ characteristic of $AdS$ solutions. The second strategy, which we will present in Appendix~\ref{app:BPStrick}, makes clever use of the relations between the asymptotics of BPS and almost-BPS solutions and turns out to give exactly the same class of solutions 
as the first strategy.

\subsection{Solutions with NHEK infrared}\label{ssec:2center}
The supergravity solutions of~\S~\ref{subsec:completion} can have multiple centers where the geometry near one of the centers is NHEK.
Our starting point to look for solutions with $AdS$ asymptotics is a two-center solution where in addition to the non-BPS D1-D5-P-KK black hole that becomes the NHEK geometry in the infrared we add another smooth center corresponding to a supertube. The asymptotics of this solution are prototypical for all solutions in the class constructed in~\cite{Bena:2012wc} and, as we will see, the conditions that  ensure that the final solution is asymptotically-$AdS_3$ do not depend on the particular distribution of centers and charges in the infrared.

One can ask whether there is any reason behind our strategy to try to embed the NHEK solution in an asymptotically-$AdS_3$ solution that has two or more centers, and hence topological-nontrivial three-cycles. Our original inspiration came from three-charge BPS black ring solutions \cite{Bena:2004de,Elvang:2004ds,Gauntlett:2004qy} embedded in an 
 asymptotically $AdS_3 \times S^3$ solution \cite{Bena:2004tk}: These rings have another $AdS_3$ region in the vicinity of the black ring center, with smaller $AdS$ radius; thus the black ring solution can be thought of as an RG flow from a CFT in the UV to a CFT with lower central charge in the IR. The full solution has a topologically-nontrivial three-sphere at whose North Pole this smaller $AdS_3$ sits.  
 
 One can also see by direct calculation that a multicenter solution is necessary to get a NHEK region in the infrared. One could try to start from a single-center solution and play with the constants in the harmonic functions to obtain a cohomogeneity-one solution with an $AdS_3$ UV, but one will find that the infrared of this solution is always $AdS_3$. This is essentially because in a single-center solution the $K_I$ harmonic functions are constant\footnote{In an almost-BPS solutions the $K_I$ functions cannot have poles where the Taub-NUT harmonic function $V$ has poles, as these give rise to singularities or diverging moduli~\cite{Bena:2012ub}.}, and hence their value is the same at infinity and at the black hole. The only way to get something other than $AdS_3$ in the infrared is to makes those values different by introducing additional sources for $K_I$, and thus additional centers. 
 
Hence, we begin by considering a two-center solution with a Taub-NUT base containing a non-BPS black hole and a supertube at distance $R$ in the ${\mathbb R}^3$ base of the Taub-NUT space. The solution is easiest to describe using two sets of spherical coordinates, one centered at the black hole $(\rho,\theta)$ and another centered at the supertube $(\Sigma,\theta_\Sigma)$, related by
\begin{equation}
\Sigma=\sqrt{\rho^2+R^2-2\rho R \cos\theta}\,, \qquad \cos\theta_\Sigma = \frac{\rho \cos\theta-R}{\Sigma}\,,\qquad  \cos\theta'_\Sigma=\frac{\rho-R\cos\theta}{\Sigma}\,.
\end{equation}
The metric and three-form flux of the solution are given by~\eqref{eq:widetildeds} and~\eqref{eq:widetildeF3} of~\S~\ref{subsec:completion} and are specified by the functions $V,K_I,L_I,M,\mu$ and the one-forms $w_I,\omega,v_I,\nu$.~\footnote{In Section~6 of~\cite{Bena:2009ev} the solution for a non-BPS black ring and a black hole in Taub-NUT was found in the type IIA duality frame. We adapt this solution for our purpose, by turning off a charge and two dipole charges of the black ring to transform it into a supertube.}
The functions $V,K_I,L_I,M$ are 
 \begin{align}\label{eq:harmGold}
 \begin{split}
V &= v_{0}+\frac{\qKK}{\rho}\,, \qquad M=m_{0}+\frac{J}{\rho}+\frac{j}{\Sigma} +c\frac{\cos\theta}{\rho^2} 
\,,\\
K_I &= k_I^{0} +\frac{d_I}{\Sigma}\,, \qquad L_I = \ell_I^{0}+\frac{Q_I}{\rho}+\frac{e_I}{\Sigma}\,.
\end{split}
\end{align}
The supertube charges, dipole charges and angular momentum are denoted by $\{e_I\}=\{e_1,e_2,0\}$, $\{d_I\}=\{0,0,d_3\}$ and $j=\frac{e_1 e_2}{2d_3}$, and the black hole has charges $\{Q_I\}=\{Q_1,Q_2,Q_3\}$ and angular momentum $J$. The KK gauge potential is $A=- \qKK \cos\theta d\phi$.
The warp factors are given by $Z_I=L_I$. The angular momentum one-form $k=\mu(d\psi+A)+ \omega$ is given by\footnote{Note that in the expression for $\mu$ given in~\cite{Bena:2009ev} there is a typo: the factor of $\cos\theta$ in the last term in~\eqref{eq:mu} is missing.}
\begin{eqnarray}
\label{eq:mu}
 \mu &=& \frac{M}{V} + \frac{\ell_3^0 d_3}{2\Sigma} + \frac{Q_3 d_3 v_0}{2V \rho \Sigma} + \frac{Q_3 d_3 \qKK}{R} \frac{\cos\theta}{V \rho \Sigma}  
 \,,\\
\label{eq:omega}
 \omega &=& \Big\{\kappa + J \cos\theta + j\cos\theta_\Sigma - c\frac{\sin^2\theta}{\rho} - \frac{Q_3 d_3 \qKK}{R}\frac{\sin^2\theta}{\Sigma}  
 \\
 &&-\frac{\ell_3^0 d_3 v_0}{2} \cos\theta_\Sigma - \frac{(\qKK \ell_3^0 +v_0 Q_3)d_3}{2R} \cos\theta'_\Sigma  
 \Big\} d\phi\,.\nn
\end{eqnarray}
The magnetic potentials $a_I=K_I (d\psi +A) + w_I$ are specified by
\begin{align}
\begin{split}
&\quad \quad w_{1} = k_{1}^0 \qKK \cos\theta d\phi\,,\quad w_{2} = k_{2}^0 \qKK \cos\theta d\phi\,,\\
&w_3 = \left(k_3^0 \qKK \cos\theta - d_3 v_0\cos\theta_\Sigma - d_3\qKK\cos\theta'_\Sigma/R \right) d\phi \,.
\end{split}
\end{align}
and the one-forms $\nu,v_I$ in~\eqref{eq:tildeoneforms} are
\begin{align}
\begin{split}
\nu=& \Big[ (-k_1^0 Q_1 - k_2^0 Q_2- k_3^0 Q_3 - k_1^0 k_2^0 k_3^0 \qKK ) \cos\theta + \\
 &(-k_1^0 e_1 - k_2^0 e_2 + k_1^0 k_2^0 d_3 v_0 + \ell_3^0 d_3 ) \cos\theta_\Sigma \\
 &+ (Q_3 + k_1^0 k_2^0 \qKK) d_3\cos\theta'_\Sigma/R \Big] d\phi\,,
 \end{split}
\end{align}
and
\begin{align}
\begin{split}
 &v_{1}= (Q_{1} \cos\theta + e_{1} \cos\theta_\Sigma) d\phi+k_{2}^0 w_3\,,\quad 
v_{2} = (Q_{2} \cos\theta + e_{2} \cos\theta_\Sigma )d\phi+k_{1}^0 w_3\,,\\
& \qquad \qquad \qquad \qquad \qquad v_3 = (Q_3 +k_1^0 k_2^0 \qKK) \cos\theta d\phi\,.
\end{split}
\end{align}

Regularity requires the absence of Dirac strings at the poles $\theta=0,\pi$ of the two-sphere which implies that $\omega$ must vanish there:
 \begin{eqnarray}
  0=\omega \Big|_{\theta=0} &=&\kappa +J +s_- j - s_- \frac{\ell_3^0 d_3}{2} \left( v_0 +\frac{\qKK}{R}\right) - s_- \frac{Q_3 d_3 v_0}{2R} 
  \,,\\
   0=\omega \Big|_{\theta=\pi} &=&\kappa -J -s_+ j + s_+ \frac{\ell_3^0 d_3}{2} \left( v_0 -\frac{\qKK}{R}\right) - s_+ \frac{Q_3 d_3 v_0}{2R} 
   \,,
 \end{eqnarray}
 with $s_\pm = {\rm sign}(r\pm R)$. 
This gives three conditions which determine $\kappa, J, j$:
\begin{eqnarray}
 J&=&- \kappa = -\frac{(\ell_3^0 \qKK - v_0 Q_3) d_3}{2R}\,,\\
 j&=&\frac{\ell_3^0 d_3}{2} \left(v_0+\frac{\qKK}{R}  \right) + \frac{Q_3 d_3 v_0}{2R}
 \,.\label{eq:bubble}
\end{eqnarray}
The second equation is interpreted as the bubble equation that determines the distance $R$ between the centers in the $\mathbb{R}^3$ base of the Taub-NUT space, or the radius of the supertube in the five-dimensional solution.

\subsection{Hunting for AdS asymptotics}\label{ssec:rules}

We now explore the UV structure of the metric~\eqref{eq:widetildeds} specified by the functions in~\S~\ref{ssec:2center}. To simplify the analysis we make use of the fact that the third spectral flow corresponds to a coordinate transformation and so we can set $\g_3=0$ without loss of generality. On the other hand, for the IR to correspond to NHEK, $\g_1$ and $\g_2$ must be non-vanishing.
For the UV to be $AdS$ we need the radial part of the metric to behave as $g_{\rho \rho} \sim d\rho^2/\rho^2$ at large $\rho$. 
From~\eqref{eq:widetildeds} and \eqref{eq:widetildeds4} we read off the radial part of the metric:
\begin{equation}\label{eq:grhorhogeneral}
 g_{\rho \rho} = \widetilde V \sqrt{\widetilde Z_1 \widetilde Z_2}= \sqrt{N_1 N_2}  \,,
\end{equation}
where we made use of~\eqref{eq:tildefunctions}.
The large-$\rho$ expansion of the $N_I$ gives
\begin{equation}
 N_I = n_I + \pi_I/\rho+\sigma_I/\rho^2 + \mathcal{O}(1/\rho^3)\,,
\end{equation}
with
\begin{eqnarray}
 n_1&=&\ t_1^2 \ell_1^0 v_0 + \g_1 \left(\g_1 \ell_2^0 \ell_3^0 - 2t_1 m_0\right)\,,\\
 \pi_1&=&   \g_1^2 (\ell_3^0 (Q_2+e_2)+\ell_2^0 Q_3) + t_1^2\left((Q_1+e_1)v_0 + \ell_1^0 \qKK\right) - 2 \g_1 t_1  v_0 \ell_3^0 d_3\,,\\
  \sigma_1&=& t_1^2 (Q_1+e_1) \qKK + \g_1^2 (Q_2+e_2) Q_3 - \g_1 t_1 (\ell_3^0 \qKK + v_0 Q_3) d_3  \label{eq:sigma}\\
  & &- t_1 \g_1 \left[ 2 c -R\left(\frac{t_1}{\g_1} e_1 v_0+\frac{\gamma_1}{t_1} e_2 l_3^0\right) +  d_3 \left(l_3^0 (\qKK+2 v_0 R)+ Q_3 (2 \qKK/R+ v_0)\right)\right] \cos\theta\,,\nn 
\end{eqnarray}
and similar expressions for $n_2,\pi_2,\sigma_2$ (obtained by exchanging $1 \leftrightarrow 2$), where $t_I=1+k_I^0 \gamma_I$ is defined by expanding the function $T_I = t_I + \mathcal{O}(1/\rho)$.
To have the constant and $1/\rho$ terms in~\eqref{eq:grhorhogeneral} vanish in the large $\rho$ expansion requires
\begin{equation}\label{eq:npi}
 n_1=n_2=\pi_1=\pi_2=0\,.
\end{equation}
After imposing these constraints the leading-order term in the radial part of the metric becomes
\begin{equation}
 g_{\rho \rho} \approx \sqrt{\sigma_1 \sigma_2}/\rho^2\,.
\end{equation}
In general this expression contains terms proportional to $\cos\theta$ that would give rise to warped $AdS$ asymptotics, and since our purpose is to obtain ``normal''  $AdS$ we require these terms to vanish. To leading order this can be achieved by writing $\sigma_I=\sigma_I^0 + \sigma_I^\theta \cos\theta$, with $\sigma_1^0$ and $\sigma_1^\theta$ corresponding respectively to the first and second line of~\eqref{eq:sigma} (and similarly for $\sigma_2$), and imposing
\begin{equation}\label{eq:costheta}
 \sigma_1^\theta = \sigma_2^\theta =0\,.
\end{equation}
The leading radial part of the metric is then simply given by
\begin{equation}\label{eq:grhorho}
 g_{\rho \rho}\approx \sqrt{\sigma_1^0 \sigma_2^0}/\rho^2\,,
\end{equation}
subject to relations between the parameters of the solution following from the constraints~\eqref{eq:npi} and~\eqref{eq:costheta}.
Regularity imposes further constraints: for absence of closed timelike curves (CTCs) we need
\begin{equation}\label{eq:CTCZ123}
Z_I V \geqslant 0\,,                                                                                                                                                                                                      \end{equation}
as well as
\begin{eqnarray}\label{eq:CTC}
& Z_1 Z_2 Z_3 V - \mu^2 V^2 -\omega^2 \geqslant 0\,,&\\
 &\Downarrow&\nn \\ \label{eq:CTC012}
& (CTC)_0+(CTC)_1/\rho+(CTC)_2/\rho^2+\dots \geqslant 0&\,,
\end{eqnarray}
where in the second line we have expanded the no-CTC condition in the UV. Since the constants $\ell_I^0$ are non-negative and the charges $Q_I$, $e_1$, $e_2$ in $Z_I$ are strictly positive, to avoid CTCs we also need the constant $v_0$ to be non-negative and the KK monopole charge $\qKK$ to be strictly positive. Altogether we have:
\begin{equation}\label{eq:constcharg}
 \ell_I^0, v_0 \geqslant 0\,, \quad Q_I, e_1, e_2, \qKK > 0\,.
\end{equation}
The condition~\eqref{eq:CTC} is equivalent to the existence of a global time function \cite{Berglund:2005vb}.

\subsection{Constraints for AdS asymptotics}\label{ssec:solution}
In the following we will solve the constraints~\eqref{eq:npi} and~\eqref{eq:costheta} of~\S~\ref{ssec:rules}. It is useful to consider separately the solutions where $\ell_3^0 =0$ and $\ell_3^0 \neq 0$. 

\subsubsection{No Solution for $\ell_3^0\not =0$}
We will first try to find an $AdS_3$ throat with $\ell_3^0\not =0$. 
To solve the constraints~\eqref{eq:npi} we need to impose
\begin{equation}
t_1 \not = 0 \quad \text{and}\quad t_2\not = 0\,. 
\end{equation}
One can then solve $n_1=0$ to obtain
\begin{equation}
 m_0=\frac{1}{2}\left[\frac{t_1}{\g_1}\ell_1^0 v_0+\frac{\g_1}{t_1}\ell_2^0 \ell_3^0\right]\,.
\end{equation}
The second constraint $n_2=0$ is solved by
\begin{equation}
 v_0=\ell_3^0 \frac{\g_1 \g_2}{t_1 t_2}\quad \text{or} \quad \ell_2^0 = \frac{\g_2 t_1}{\g_1 t_2}\ell_1^0\,.
\end{equation}
In principle we have to consider both solutions. However, it turns out that canceling the negative leading-order contribution $(CTC)_0 = -m_0^2+\ell_1^0 \ell_2^0 \ell_3^0 v_0$ in~\eqref{eq:CTC012} both conditions have to be imposed:
\begin{equation}\label{eq:v0m0rel}
 \left[n_1=n_2=(CTC)_0=0\right]~~\Leftrightarrow~~\left[ v_0=\ell_3^0 \frac{\g_1 \g_2}{t_1 t_2}\text{~,~~~} \ell_2^0 = \frac{\g_2 t_1}{\g_1 t_2}\ell_1^0\quad \text{and}\quad m_0=\ell_1^0 \ell_3^0 \frac{\g_2}{t_2}\right]\,.
\end{equation}
Further solving $\pi_1=0$ for $d_3$ gives
\begin{equation}\label{eq:d3rel}
 d_3=\frac{1}{2\ell_3^0}\left[\frac{t_1}{\g_1}(Q_1+e_1)+\frac{t_2}{\g_2}(Q_2+e_2)+\frac{\ell_1^0}{\ell_3^0} \frac{t_1t_2}{\g_1\g_2} \left(\frac{t_1}{\g_1}  \qKK+\frac{\g_2}{t_2} Q_3\right)\right]\,.
\end{equation}
With this, the constraint $\pi_2=0$ is automatically satisfied. For the metric to correspond to unwarped $AdS$ at large $\rho$ we have to impose~\eqref{eq:costheta} which can be achieved by setting
\begin{eqnarray}\label{eq:crel}
 c&=&-\Big\{\frac{R}{2}\left[\ell_3^0 \left(\frac{\g_2}{t_2} Q_1+ \frac{\g_1}{t_1} Q_2 \right) + \ell_1^0  \left(\frac{t_1}{\g_1} \qKK+\frac{\g_2}{t_2} Q_3\right)\right]\nn \\
 &&+\frac{1}{4\ell_3^0} \frac{t_1 t_2}{\g_1 \g_2} \left[\ell_3^0\left(\frac{\g_2}{t_2} (Q_1+e_1)+ \frac{\g_1}{t_1} (Q_2+e_2)\right)+\ell_1^0 \left( \frac{ t_1}{ \g_1} \qKK+\frac{\g_2}{t_2} Q_3 \right)\right]\cdot\\
 &&\qquad \qquad \cdot \left(\frac{\g_1}{t_1}\left(\frac{t_1}{\g_1}\qKK+\frac{\g_2}{t_2} Q_3 \right)+\frac{2  Q_3 \qKK }{R{\ell_3^0}}\right)\Big\}\,.\nonumber
\end{eqnarray}
For the solution to be free of CTCs we need to impose~\eqref{eq:CTC} subject to the constraints~\eqref{eq:v0m0rel},~\eqref{eq:d3rel} and~\eqref{eq:crel}. The leading non-vanishing term is $(CTC)_2/\rho^2$ which, evaluated at $\theta=0$ or $\theta=\pi$, is:
\begin{align}
\begin{split}
 0 \leqslant (CTC)_2 = &-\frac{1}{4} \left[ (\ell_3^0)^2\left(\frac{\g_2}{t_2} (Q_1+e_1)-\frac{\g_1}{ t_1}(Q_2 +e_2)\right)^2\right.\\
 &+\left.(\ell_1^0)^2 \left(\frac{t_1}{\g_1}\qKK+ \frac{\g_2}{t_2}  Q_3 \right)^2 +2 (\ell_1^0)^2 \left( \frac{t_1^2}{\g_1^2} \qKK^2  +\frac{\g_2^2}{t_2^2}  Q_3^2\right)\right] \,.
 \end{split}
\end{align}
This condition can only be satisfied if $(Q_1+e_1) \g_2t_1 = (Q_2+e_2) \g_1t_2$ and $\ell_1^0=0$. This, however, kills the sought after $d\rho^2/\rho^2$ term in the metric. 
Hence, the solutions in~\S~\ref{ssec:2center} with $\ell_3^0 \neq 0$ do not have an asymptotic $AdS$ metric.

\subsubsection{Solutions with $\ell_3^0  =0$}\label{TheSolution}
When $\ell_3^0$ vanishes, the non-negativity of the leading part of the no-CTC condition \eqref{eq:CTC} imposes:
\begin{equation}\label{eq:CTCl30}
 \left[(CTC)_0= -m_0^2\geqslant 0\right]\quad \Leftrightarrow\quad m_0=0\,.
\end{equation}
From the constraints $\pi_1=\pi_2=0$ one can see that one must also have $\ell_1^0=\ell_2^0=0$. 
The constraints $n_1=n_2=0$ are then automatically satisfied. 
To satisfy the no-CTC condition~\eqref{eq:CTC} at order $1/\rho^2$ implies
\begin{equation}
 \left[(CTC)_2\equiv -\left(c+Q_3 d_3(\qKK/R+v_0/2) \right)^2 \sin^4\theta \geqslant 0\right]\quad \Leftrightarrow \quad c=-\frac{d_3Q_3 (R v_0+2\qKK)}{2R}\,.
\end{equation}
Finally, the constraints $\pi_1=\pi_2=0$ are satisfied by imposing\footnote{One can also satisfy $\pi_1=\pi_2=0$ by setting $v_0=0$. This would, however, set $j=0$ which is inconsistent since the supertube has non-vanishing angular momentum.}
\begin{equation}
t_I=0 \quad \Leftrightarrow \quad k_I^0=-1/\g_I \qquad \text{for} \qquad I=1,2\,.
\end{equation}
Notice that this implies that $T_1=T_2=0$\footnote{This is a consequence of having only one non-vanishing dipole charge ($d_3$ in $K_3$). Replacing the supertube by a black ring turns on extra dipole charges, $d_1$ and $d_2$, in respectively $K_1$ and $K_2$ and in such a solution $T_1\neq 0,\,T_2\neq0$.}. With these constraints the bubble equation~\eqref{eq:bubble} becomes
\begin{equation}
 \frac{Q_3 d_3 v_0}{2R} =j \equiv  \frac{e_1 e_2}{2 d_3} \qquad \Rightarrow \qquad R=\frac{Q_3 d_3^2}{e_1 e_2} v_0\,,
\end{equation}
where we used the usual relation between the supertube charges, dipole charges and angular momentum. The distance between the centers, $R$, is determined once the constants controlling the UV are fixed.
With these constraints the leading radial metric component is $d\rho^2/\rho^2$ and the ultraviolet is free of CTCs. To ensure that there are no CTCs in the infrared we expand~\eqref{eq:CTC} for small $\rho$:
\begin{equation}
 0 \leqslant Q_1 Q_2 Q_3 \qKK - c^2 \cos^2\theta\,,
\end{equation}
which can always be satisfied by taking $Q_1,Q_2$ sufficiently large. The no-CTC conditions~\eqref{eq:CTCZ123} are automatically satisfied near the black hole. It is also trivial to check that the determinant of the UV metric at leading order is constant and negative:
\begin{equation}
 {\rm Det}(g_{UV}) \approx -\frac{(\g_1^2(Q_2+e_2))^3}{\g_2^2(Q_1+e_1)} Q_3^2 \sin^2\theta < 0\,.
\end{equation}

\subsection{Asymptotically-$AdS_3$ solutions}

We now summarize the features of the solutions that contain a NHEK region \cite{Bena:2012wc} and have an $AdS_3$ UV.
The complete bulk metric is given by~\eqref{eq:widetildeds} of~\S~\ref{subsec:completion} and is specified by the harmonic functions~\eqref{eq:harmGold} of~\S~\ref{ssec:2center}. For the UV of the metric~\eqref{eq:widetildeds} to have a leading radial component $d\rho^2/\rho^2$ we have to constrain the constants that appear in the harmonic functions that determine the solution:
\begin{equation} \label{eq:solconstraints}
m_0=0\,, \quad \ell_I^0=0\,, \quad k_1^0=-1/\g_1\,, \quad k_2^0=-1/\g_2\,.
\end{equation}
Absence of CTCs requires positivity of the charges~\eqref{eq:constcharg} and
\begin{equation}\label{eq:cconstraints}
c=-\frac{d_3Q_3 (R v_0+2\qKK)}{2R}\,.
\end{equation}

This equation enforces the cancelation of the $\sin^2 \theta\over \rho$ term in the asymptotics of $\omega$, and is necessary for avoiding CTCs only when the asymptotics is $AdS$. This cancelation is equivalent to the vanishing of the total four-dimensional angular momentum ($J_R$) of the solution. In the particular solution we consider, this angular momentum has both a contribution from the black hole as well as a contribution coming from the interaction between the magnetic dipole charge of the supertube and the electric charge of the black hole, and equation~\eqref{eq:cconstraints} forces these contributions to cancel each other.

For a more general multicenter solution containing one or more black holes (whose near-horizon regions are transformed by the generalized spectral flows into a NHEK  region), there will be more contributions to $J_{R}$ (see for example \cite{Bossard:2012ge}), and we expect that the only requirements to have asymptotically-$AdS_3$ solutions are that the constants entering in the harmonic functions satisfy~\eqref{eq:solconstraints} and that the total $J_{R}$ vanishes.

Of course, the more general multicenter solutions will also have to satisfy the corresponding bubble equations, which for the two-center solution we focus on are quite simple:
\begin{equation}\label{eq:bubblefinal}
R=\frac{Q_3 d_3^2}{e_1 e_2}v_0\,,
\end{equation}
but in general will be much more complicated~\cite{Bena:2009ev}.


In terms of a more standard radial coordinate $r^2=4\qKK \rho$ the UV metric at leading order is 
  \begin{align} \label{eq:UVmetric}
   ds^2_{UV}&\approx \frac{4\lUV^2}{r^2}dr^2 + \frac{r^2}{4\lUV^2} (-dT^2 + dY^2)+ \lUV^2 \left[(d\theta^2+\sin^2\theta d\phi^2)+ \left(\frac{d\psi}{N}-{\rm sgn} (\g_1 \g_1)\cos\theta d\phi\right)^2\right]\nn\\
   &+\frac{1}{\lUV^2} \Big[ \alpha(-dT+dY)^2 -\beta N (dT+3 dY)\left(\frac{d\psi}{N}-{\rm sgn} (\g_1 \g_1) \cos\theta d\phi\right) \\
   &\qquad \quad - \gamma \cos\theta (-dT^2+dY^2)\Big]+ \tau_{\rm UV} \, ds_{T^4}^2 \,,\nn
  \end{align}
where
\begin{equation}
 T = \sqrt{\frac{Q_3 v_0 }{\qKK}} \frac{\g_1 \g_2}{v_0} t\,, \qquad Y=\sqrt{\frac{Q_3 v_0}{\qKK}} \left(\frac{\g_1 \g_2}{v_0} t+z-k_3^0 \psi+\frac{d_3}{R}N {\rm sgn} (\g_1 \g_2)\phi\right)\,,
\end{equation}
and we defined
\begin{align}
&\qquad \qquad \qquad \qquad  N=|\g_1 \g_2| Q_3\,, \quad \lUV^2=N \sqrt{(Q_1+e_1)(Q_2+e_2)}\,,\\
&\alpha=\frac{\qKK^2}{v_0} \,, \quad 
\beta=d_3 \sqrt{Q_3 \qKK v_0}\,,\quad \gamma=\qKK \left(\frac{e_1}{Q_1+e_1}+\frac{e_2}{Q_2+e_2}\right) \frac{R}{2}\,, \quad \tau_{\rm UV} = \left|\frac{\g_1}{\g_2}\right| \sqrt{\frac{Q_2+e_2}{Q_1+e_1}}\,.\nn
\end{align}
The term in~\eqref{eq:UVmetric} proportional to $\beta$ corresponds to a spectral flow and so we can remove it by a coordinate transformation: 
\begin{equation}
  \Psi = \psi - \frac{\beta N^2}{2\lUV^4} (T+3Y)\,,
\end{equation}
yielding
\begin{eqnarray} \label{eq:UVmetricfinal}
 ds^2_{UV}&\approx& \frac{4\lUV^2}{r^2}dr^2 + \frac{r^2}{4\lUV^2} (-dT^2 + dY^2)+ \lUV^2 \left[(d\theta^2+\sin^2\theta d\phi^2)+ \left(\frac{d\Psi}{N}-{\rm sgn} (\g_1 \g_2)\cos\theta d\phi\right)^2\right]\nn\\
&&+ \alpha'(-dT+dY)^2  -2 \beta' dY (dT+dY) - \gamma' \cos\theta (-dT^2+dY^2)+ \tau_{\rm UV}\, ds_{T^4}^2\,,
\end{eqnarray}
where
\begin{equation}
 \alpha'=\frac{1}{\lUV^2}\left(\alpha-\frac{\beta^2 N^2}{4 \lUV^4}\right)\,, \quad \beta'=\frac{\beta^2 N^2}{\lUV^6}\,, \quad \gamma'=\frac{\gamma}{\lUV^2}\,.
\end{equation}
The leading part of the UV metric~\eqref{eq:UVmetricfinal} is $AdS_3 \times S^3/{\mathbb{Z}_N\times T^4}$ with subleading perturbations that trigger the RG flow to NHEK. 

This establishes that all the geometries containing NHEK regions in the infrared~\cite{Bena:2012wc} can be embedded 
into asymptotically-$AdS_3$ solutions, and hence that the theories dual to NHEK can arise as fixed points of Renormalization Group flow from the D1-D5 CFT dual to $AdS_3$. Different multicenter solutions will correspond to different RG flows, and we leave the detailed study of these flows for future work.

\section*{Acknowledgements}
We would like to thank Monica Guic\u a for valuable discussions and collaboration in the early state of this project and Geoffrey Comp\`{e}re and Diego Hofman for interesting discussions and comments on the manuscript.
The work of I.B. was supported in part by the ERC Starting Grant 240210 String-QCD-BH, by the John Templeton Foundation Grant 48222 and by a grant from the Foundational Questions Institute (FQXi) Fund, a donor advised fund of the Silicon Valley Community Foundation on the basis of proposal FQXi-RFP3-1321 (this grant was administered by Theiss Research).
The work of A.P. is supported by National Science Foundation Grant No. PHY12-05500.
I.B. and A.P. are grateful to the Centro de Ciencias de Benasque Pedro Pascual and to the Aspen Center for Physics for hospitality and support via the National Science Foundation Grant No. PHYS-1066293.

\appendix{

\section{Throats with a NHEK - an alternative approach}\label{app:BPStrick}

In~\S~\ref{sec:UV} we carried out a systematic analysis for embedding the multicenter solutions containing NHEK  regions in an asymptotic $AdS$ geometry and identified a set of constraints that have to be satisfied. We now describe a second approach that yields the same set of constraints.

The leading and subleading asymptotics of an almost-BPS solution is controlled by 17 parameters, and the same is true for its BPS equivalent. Generalized spectral flows shuffle these parameters in a certain way, and at the end one has to fix 7 of these parameters in order to obtain an $AdS_3$ UV.  Since the near-horizon of any almost-BPS black hole can be transformed into a NHEK geometry by choosing the appropriate generalized spectral flow parameters~\cite{Bena:2012wc}, this implies that there should be at least a 10-parameter family of solutions with NHEK IR and an $AdS_3$ UV. However, as emphasized in the Introduction, most of the solutions have closed timelike curves, which have to be eliminated (as we also did in the approach described in~\S~\ref{sec:UV}). At the end we will find that this approach only produces {\it one} way to get a solution that is asymptotically $AdS_3$ and free of CTCs and this is exactly the same solution as the one obtained in~\S~\ref{sec:UV} !

\paragraph{The method.}
The procedure which we will now develop makes clever use of the fact that the only difference between the asymptotics of BPS and almost-BPS solutions is the sign of one of the electric fields~\cite{Goldstein:2008fq}, of the fact that BPS solutions transform under generalized spectral  flows into other BPS solutions~\cite{Bena:2008wt}, and of the fact that BPS solutions develop an $AdS_3$ UV region when certain of their moduli are put to zero\footnote{As in~\S~\ref{sec:UV}, without loss of generality, we set to zero the third spectral flow.}. The procedure we follow is summarized in this illustration:
{\renewcommand{\arraystretch}{1.5}
\renewcommand{\tabcolsep}{0.2cm}
\begin{center}
\begin{tabular}{lll}
{\bf BPS:} $AdS_3$ UV &\hspace{1.5cm}&$\;$ {\bf non-BPS:} NHEK IR \underline{\it and} $AdS_3$ UV ?\\
$\downarrow$ {\small \it 2 BPS gSFs}&&$\;$ $\uparrow$ {\small \it 2 non-BPS gSFs}\\
{\bf BPS}$_+$ & $\;$ $\stackrel{\text{UV}}{\equiv}$ $\;$ & $\;$ {\bf almost-BPS}$_-$
\end{tabular}
\end{center}}
We start with a BPS solution that has an asymptotic $AdS_3$ geometry. We then perform two BPS generalized spectral flow transformations (gSF) that yields another BPS solution whose asymptotics we identify with those of the almost-BPS solution of~\S~\ref{ssec:2center} (to distinguish the BPS and almost-BPS solutions whose asymptotics we are matching we use, respectively, $+$ and $-$ sub/super scripts). This fixes some of the moduli of the almost-BPS solution. Applying two non-BPS generalized spectral flow transformations to this almost-BPS solution yields a non-BPS solution which contains a NHEK geometry in the IR and whose UV we show to correspond to an $AdS_3$ geometry. 

\paragraph{Hunting for $AdS_3$.}
In a BPS solution that is determined by 8 harmonic functions \cite{Bates:2003vx,Gauntlett:2004qy,Bena:2005ni} one can read off the asymptotics and charges from $H=h+\sum_i \frac{\Gamma_i}{\rho_i}$, where $h$ is the vector of constants and $\Gamma_{\rm tot}=\lim_{\rho \to \infty} \sum_i \frac{\Gamma_i}{\rho_i}$ is the total charge vector as determined by the harmonic functions $H=(V,\{K_I\},\{L_I\},M)$:
\begin{equation}
 h=(v_{0},\{k_I^{0}\},\{\ell_I^{0}\},m_{0})\,, \quad \Gamma_{\rm tot}= (v_{\rm tot},\{k_I^{\rm tot}\},\{\ell_I^{\rm tot}\},m_{\rm tot})\,,
\end{equation} 
A BPS solution has $AdS_3$ asymptotics if the vector of constants takes the form~\cite{Bena:2011zw}:
\begin{equation}
  h_{AdS}=\left(0,\{0,0,0\},\{0,0,{\rm l}_3^0\},{\rm m}_0\right)\,,
  \label{eq:hBPSAdS}
\end{equation}
with ${\rm l}_3^{0}\neq 0$. The requirement that\footnote{The symplectic product is given by $
 \langle \Gamma_{\rm tot}, h \rangle = 2(v_{\rm tot} m_{0} - m_{\rm tot} v_{0})+(k_I^{\rm tot} \ell_I^{0}- \ell_I^{\rm tot} k_I^{0})\,.$} $\langle \Gamma_{\rm tot},h_{AdS}\rangle=0$ fixes ${\rm m}_0=-{\rm k}_3^{\rm tot}/{\rm v}_{\rm tot}$ with ${\rm v}_{\rm tot}\neq 0$.
Performing two BPS spectral flow transformations with parameters $\gamma_1$ and $\gamma_2$,
 corresponds to the supergravity transformations~\cite{Bena:2008wt}\footnote{Note that work in the conventions of~\cite{Dall'Agata:2010dy} which have an extra factor of 2 multiplying $M_+$ when compared to the conventions of~\cite{Bena:2008wt}.}:
 \begin{align}\label{flow_trans}
 \begin{split}
M &\rightarrow M_+=M\,, \\
L_I &\rightarrow L_I^+=L_I - 2\gamma_I M\,, \\
K_I &\rightarrow K_I^+=K_I - C_{IJK} \gamma_J L_K  + C_{IJK} \gamma_J
\gamma_K M\,, \\
V &\rightarrow V_+=V +  \gamma_I K_I - \frac{1}{2} C_{IJK} \gamma_I \gamma_J L_K +
\frac{1}{3} C_{IJK} \gamma_I \gamma_J \gamma_K M \,.
\end{split}
\end{align}
This yields another BPS solution with the new asymptotics:
\begin{equation}
h_{2gSF\, AdS}=\left(-\gamma_1 \gamma_2 {\rm l}_3^{0},\left\{-\gamma_2 {\rm l}_3^{0},-\gamma_1 {\rm l}_3^{0} ,2\gamma_1 \gamma_2 {\rm m}_{0}\right\},\left\{-2\gamma_1 {\rm m}_{0},-2\gamma_2 {\rm m}_{0}, {\rm l}_3^{0}\right\},{\rm m}_{0}\right)\,.
\label{eq:hBPS}
\end{equation}
For~\eqref{eq:hBPS} to correspond to the asymptotics of the almost-BPS solution of~\S~\ref{ssec:2center} we have to identify up to order $1/\rho$:
\begin{equation}
 V_+ \equiv V_-\,,  \qquad Z_I^+\equiv Z_I^-\,, \qquad (a_I^+ -k_+)\equiv-(a_I^- -k_-)\,, \qquad k_+\equiv k_-\,.
 \label{eq:ID1}
\end{equation}
The gauge potentials, warp factors and angular momentum of the almost-BPS solution are given in~\S~\ref{ssec:2center}. Those of the BPS solution are given by
\begin{eqnarray}
a_I^+&=&\frac{K_{I}^+}{V_+} (d\psi+A_+)+w_I^+\,,\\
Z_I^+&=&L_I^+ + \frac{C_{IJK}}{2} \frac{K_J^+ K_K^+}{V_+}\,,\\
\mu_+&=&M_+ +\frac{1}{2}\frac{K_I^+ L_I^+}{V_+} + \frac{C_{IJK}}{6} \frac{K_I^+ K_J^+ K_K^+}{V_+^2}\,,
\end{eqnarray}
with the constants in the harmonic functions $H_+=\{V_+,\{K_I^+\},\{L_I^+\},M_+\}$ specified by~\eqref{eq:hBPS}.
The identifications~\eqref{eq:ID1} then become~\cite{Bena:2013gma}
\begin{equation}
 V_+ \equiv V_-\,, \qquad Z_I^+ \equiv Z_I^-\,,\qquad \mu_+ \equiv \mu_-\,,\qquad \left(\frac{K_I^+}{V_+}-\frac{\mu_+}{Z_I^+}\right) \equiv -\left(K_I^- -\frac{\mu_-}{Z_I^-}\right)\,.
\end{equation}
This yields the following relations between the BPS and almost-BPS solutions (up to order $1/\rho$):
\begin{eqnarray}
V_+&\equiv&V_-\,,\\
K_I^+&\equiv&V_-\left(-K_I^- + 2 \frac{\mu_-}{Z_I^-}\right)\,,\\
L_I^+&\equiv&L_I^- - \frac{C_{IJK}}{2} \frac{K_J^+ K_K^+}{V_+}\,,\\
M_+&\equiv&\mu_- -\frac{1}{2} \frac{K_I^+ L_I^+}{V_+}+\frac{C_{IJK}}{6} \frac{K_I^+ K_J^+ K_K^+}{V_+^2}\,,
\end{eqnarray}
from which we can read off $h_+ = (v_{0,+},\{k_I^{0,+}\},\{\ell_I^{0,+}\},m_{0,+})$:
\begin{eqnarray}
 v_{0,+}&=&v_{0,-}\,,\\
 k_I^{0,+}&=&-v_{0,-} k_I^{0,-}+\frac{2 m_{0,-}}{\ell_I^{0,-}}\,,\\
 \ell_I^{0,+}&=&\ell_I^{0,-} -\frac{C_{IJK}}{2} \frac{(\ell_J^{0,-} k_J^{0.-} v_{0,-}-2m_{0,-})(\ell_K^{0,-} k_K^{0.-} v_{0,-}-2m_{0,-})}{v_{0,-} \ell_J^{0,-} \ell_K^{0,-} }\,,\\
 m_{0,+}&=&\frac{1}{2 v_{0,-}^2 \ell_1^{0,-} \ell_2^{0,-} \ell_3^{0,-}}\Big(8 m_{0,-}^3-4 v_{0,-}  m_{0,-}^2 \sum_I k_I^{0,-}\ell_I^{0,-} + \\
 &&+v_{0,-}^2 \ell_1^{0,-} \ell_2^{0,-} \ell_3^{0,-}(\sum_I  k_I^{0,-} \ell_I^{0,-} - v_{0,-} k_1^{0,-} k_2^{0,-} k_3^{0,-})\,,\nn\\
 &&+ v_{0,-}^2 m_{0,-} C_{IJK} k_J^{0,-} \ell_J^{0,-} k_K^{0,-} \ell_K^{0,-} - 4 v_{0,-} m_{0,-} \ell_1^{0,-} \ell_2^{0,-} \ell_3^{0,-}\Big)\,.
\end{eqnarray}
From $h_+  = h_{2gSF\, AdS}$ with~\eqref{eq:hBPS} we can solve for the constants of the almost-BPS solution:
 \begin{align}
  \label{eq:hnonBPSAdS}
  \begin{split}
& v_{0,-}=-\g_1 \g_2 {\rm l}_3^{0}\,,\quad k_{1}^{0,-}=-1/\g_1\,, \quad k_{2}^{0,-}=-1/\g_2\,, \quad k_3^{0,-}=2\frac{{\rm m}_{0}}{{\rm l}_3^{0}}\,,\\
& \qquad \qquad \qquad\quad  \qquad \ell_I^{0,-}=0\,,\quad m_{0,-}=0\,.
\end{split}
 \end{align}
Note, that as in~\S~\ref{ssec:solution} we can use the bubble equation and the supertube relation between charges, dipole charge  and angular momentum to obtain the relation:
\begin{equation}
 R=\frac{Q_3 d_3^2}{e_1 e_2} v_{0,-}\,.
  \label{eq:hnonBPSAdSv}
\end{equation}
We can use this to solve for ${\rm l}_3^{0}$ in~\eqref{eq:hnonBPSAdS} and plug the result into the expression for $k_3^{0,-}$.
Since ${\rm m}_0$ is a free parameter, $k_3^{0,-}$ is in fact unconstrained. With this remark and dropping the ``$-$'' in the constraints~\eqref{eq:hnonBPSAdS} we find that the constants entering in the harmonic functions determining the solution are:
\begin{equation}
 m_0=0\,,\quad \ell_I^{0}=0\,, \quad k_{1}^{0}=-1/\g_1\,, \quad k_{2}^{0}=-1/\g_2\,,
\end{equation}
which agrees precisely with the results~\eqref{eq:solconstraints} of our analysis in~\S~\ref{ssec:solution}. The extra condition on $c$ in~\eqref{eq:cconstraints} follows from the requirement that there be no closed time-like curves - a condition we also have to impose here. Hence, the method described here yields exactly the same solution as the one discussed in~\S~\ref{ssec:solution} and we take this as a remarkable confirmation that we have really identified \emph{the} flow from $AdS_3$ to NHEK.

\section{Details of the R-R field computation} \label{app:F3}

In~\cite{Bena:2012wc} it was shown that a large class of supergravity solutions with a NHEK infrared can be obtained by dualizing the twice-spectrally-flowed almost-BPS solutions that were constructed in the M2-M2-M2 duality frame in~\cite{Dall'Agata:2010dy} to the D1-D5-P duality frame. In~\cite{Bena:2012wc} this duality transformation has been performed on the metric and dilaton, but not the R-R fields. We now complete this class of supergravity solutions by computing the R-R fields in the D1-D5-P duality frame.

\paragraph{From M2-M2-M2 to D1-D5-P.} The metric and three-form gauge potential of the twice-spectrally flowed M2-M2-M2 solution obtained in~\cite{Dall'Agata:2010dy} are given by
\begin{align}\label{eq:Mmetricflux}
\begin{split}
d s^2_{11} &= -\tilde{Z}^{-2} \left(d t + \tilde{k}\right)^2 + \tilde{Z}d s^2_4 + \sum_{I=1}^3 \frac{\tilde{Z}}{\tilde{Z}_I}d s^2_I\,,\\
\caltA_3&=\sum_{I=1}^3 \tA_I \wedge dT_I=\sum_{I=1}^3 \Big(\ta_I-\frac{dt+\tk}{\tW_I}\Big) \wedge dT_I\,,
\end{split}
\end{align}
where $dT_I$ denote the volume forms of the three two-tori $T_I^2$ and $ds_1^2=dx_{4}^2+dx_{5}^2$, $ds_2^2=dx_{6}^2+dx_{7}^2$, $ds_3^2=dx_{8}^2+dx_{9}^2$, denote the metrics on the latter. We use the shortcut notation $\tilde{Z} \equiv (\tilde{Z}_1\tilde{Z}_2\tilde{Z}_3)^{1/3}$ and all tilded quantities are defined in \eqref{eq:tildeoneforms}~-~\eqref{eq:tildefunctions}. To bring this solution to the D1-D5-P duality frame we have to perform a Kaluza-Klein reduction on one of the torus legs followed by a sequence of three T-duality transformations. Note that this parallels the chain of dualities derived in~\cite{Bena:2008dw} for dualizing BPS solutions.

\begin{itemize}
 \item {\bf KK reduction on $x_9$.}
 The Kaluza-Klein reduction of the solution~\eqref{eq:Mmetricflux} along the $x_9$ direction (renaming the remaining leg of the torus $x_{10}\equiv z$) yields\footnote{Note that the sign in the expression for $B^{(2)}$ in~\eqref{eq:KKreduc} depends on which torus leg we compactify; if we chose to reduce along $x_{10}$ rather than $x_9$ the minus sign in~\eqref{eq:KKreduc} would turn into a plus sign. Anticipating that $B^{(2)}$ will become part of the metric and three-form flux after the final T-duality along $z$~\eqref{eq:F3Tdualz} this explains the relative sign between $dz$ and, respectively, $A_3$ in~\eqref{eq:metricMtoIIB} and $\widetilde A_3$ in~\eqref{eq:widetildeds}.}
 \begin{eqnarray}\label{eq:KKreduc}
  d s^2_{10} &=& -\frac{1}{\tilde{Z}_3 \sqrt{\tilde{Z}_1\tilde{Z}_2}}\left(d t + \tilde{k}\right)^2 + \sqrt{\tilde{Z}_1\tilde{Z}_2}d s^2_4 + \sqrt{\frac{\tilde Z_2}{\tilde Z_1}}ds_1^2+\sqrt{\frac{\tilde Z_1}{\tilde Z_2}}ds_2^2+\frac{\sqrt{\tilde{Z}_1\tilde{Z}_2}}{\tilde Z_3}dz^2 \,,\nonumber\\
~&&\nonumber\\
  \Phi &=&\frac{1}{4}\log\left(\frac{\tilde Z_1\tilde Z_2}{\tilde Z_3}\right)\,,~~
  B^{(2)}=-\tilde A_3\wedge dz \,,\\
  ~&&\nonumber\\
C^{(1)}&=&0\,,~~
C^{(3)}=\tilde A_1\wedge dT_1+\tilde A_2\wedge d T_2\,,\nonumber\\
~&&\nonumber\\
F^{(4)}&=&d C^{(3)}+d B^{(2)}\wedge C^{(1)} = d  \tilde A_1 \wedge d T_1 + d \tilde A_2 \wedge d T_2\,.\nonumber
 \end{eqnarray}
 \item {\bf T-duality along $x_5$}
 \begin{eqnarray}
  d s^2_{10} &=& -\frac{1}{\sqrt{\tilde{Z}_1\tilde{Z}_2}\tilde{Z}_3}\left(d t + \tilde{k}\right)^2 + \sqrt{\tilde{Z}_1\tilde{Z}_2}d s^2_4 + \sqrt{\frac{\tilde Z_2}{\tilde Z_1}}dx_6^2+\sqrt{\frac{\tilde Z_1}{\tilde Z_2}}(ds_2^2+d x_5^2)+\frac{\sqrt{\tilde{Z}_1\tilde{Z}_2}}{\tilde Z_3}dz^2 \,,\nonumber\\
~&&\nonumber\\
  \Phi &=&\frac{1}{2}\log\left(\frac{\tilde Z_1}{\tilde Z_3}\right)\,,~~
  B^{(2)}=-\tilde A_3\wedge dz \,,\\
  ~&&\nonumber\\
C^{(2)}&=&0\,,~~C^{(4)}=\tilde A_2\wedge d x_5\wedge d x_7\wedge d x_8\,,\nonumber\\
  ~&&\nonumber\\
F^{(5)}&=& d\tA_2 \wedge d x_5\wedge d x_7\wedge d x_8+\star_{10}\left( d\tA_2 \wedge d x_5\wedge d x_7\wedge d x_8\right)\,.\nonumber
 \end{eqnarray} 
 Note that the ten-dimensional Hodge star can be expressed as
 \begin{equation}
  \star_{10}\left( d\tA_2 \wedge d x_5\wedge d x_7\wedge d x_8\right) = - \left(\frac{\tilde{Z}^{5}_2}{\tilde{Z}^2_3\tilde{Z}^3_1}\right)^{1/4} \star_5d\tA_2 \wedge d z\wedge d x_6\,,
 \end{equation}
where the five-dimensional Hodge star is with respect to the five-dimensional metric
 \begin{equation}
   d s^2_{5} = -\frac{1}{\sqrt{\tilde{Z}_1\tilde{Z}_2}\tilde{Z}_3}\left(d t + \tilde{k}\right)^2 + \sqrt{\tilde{Z}_1\tilde{Z}_2}~d s^2_4\,.
 \end{equation}
 \item {\bf T-duality along $x_6$}
 \begin{eqnarray}
  d s^2_{10} &=& -\frac{1}{\sqrt{\tilde{Z}_1\tilde{Z}_2}~\tilde{Z}_3}\left(d t + \tilde{k}\right)^2 + \sqrt{\tilde{Z}_1\tilde{Z}_2}~d s^2_4 +\frac{\sqrt{\tilde{Z}_1\tilde{Z}_2}}{\tilde Z_3}dz^2+\sqrt{\frac{\tilde Z_1}{\tilde Z_2}}(ds_1^2+d s_2^2)\,,\nonumber\\
~&&\nonumber\\
\Phi &=&\frac{1}{4}\log\left(\frac{\tilde Z_1^3}{\tilde Z_2\tilde Z_3^2}\right)\,,~~B^{(2)}=-\tilde A_3\wedge dz\,,\quad C^{(1)}=-A_1 - dt\,,\\
&&~\nonumber\\
F^{(2)}&=&- d\tA_1\,,~~F^{(4)}=- \left(\frac{\tilde{Z}^{5}_2}{\tilde{Z}^2_3\tilde{Z}^3_1}\right)^{1/4} \star_5d\tA_2\wedge d z\,.\nonumber
 \end{eqnarray}
 
 \item {\bf T-duality along $z$}
 \begin{eqnarray}\label{eq:F3Tdualz}
  d s^2_{10} &=& - \frac{1}{\tilde{Z}_3\sqrt{\tilde{Z}_1\tilde{Z}_2}}( d  t + \tilde{k})^2 + \sqrt{\tilde{Z}_1\tilde{Z}_2}  d  s_4^2 + \frac{\tilde{Z}_3}{\sqrt{\tilde{Z}_1\tilde{Z}_2}}\left( d  z-\tilde{A}_{3} \right)^2 + \sqrt{\frac{\tilde{Z}_1}{\tilde{Z}_2}}( d  s_1^2+ d  s_2^2)\,,\nonumber\\
~&&\nonumber\\
\Phi &=&\frac{1}{2}\log\left(\frac{\tilde Z_1}{\tilde Z_2}\right)\,,~~B^{(2)}=0\,,\\
&&~\nonumber\\
F^{(3)}&=&- \left(\frac{\tilde{Z}^{5}_2}{\tilde{Z}^2_3\tilde{Z}^3_1}\right)^{1/4} \star_5d\tA_2- d\tA_1 \wedge\left(d z-\tilde A_3\right)\,.\nonumber
 \end{eqnarray}
\end{itemize}

Finally the R-R three-form flux in the D1-D5-P duality frame is given by
\begin{equation}
\widetilde F^{(3)}=    -\left(\frac{\widetilde {Z}^5_2}{\widetilde {Z}^2_3\widetilde {Z}^3_1}\right)^{1/4}\star_5 d\widetilde A_2+d  \widetilde A_1 \wedge (  \widetilde A_3 -d  z ) \,.\label{F3_short}
\end{equation}
Note that this expression has the same form as the R-R three-form flux~\eqref{eq:fluxMtoIIB} of the almost-BPS solution. However,~\eqref{F3_short} is considerably more complicated. In particular, the action of the five-dimensional Hodge star on this expression involves repeated use of expressions of the tilded forms and functions, as well the simplification of the result using the almost-BPS equations~\eqref{eq:almostBPS0}~-~\eqref{eq:almostBPS4}. 

To give the explicit form we make some simplifying assumptions. We will restrict ourselves to solutions with $\mu=0$ (yet keeping $\omega\neq0$), and allow the $K_I$ to be constants ($dK_I=0$). Computing the explicit form of the three-form flux for more general solutions is much more complicated and we will not address it here. 

As mentioned before, the third spectral flow corresponds to a coordinate transformation in the D1-D5-P duality frame, so without loss of generality we only focus on the solution obtained after two spectral flows and set $\gamma_3=0$. This leads to considerable simplifications. Since $T_3=1+\gamma_3 K_3$ and $K_3={\rm const}<\infty$ we have ${T_3=1}$ and the tilded functions~\eqref{eq:tildeoneforms}~-~\eqref{eq:tildefunctions} relevant here simplify to
\begin{equation}\label{eq:tildesimp1}
 \tV=|T_1 T_2 V - \g_1 \g_2 Z_3|\,, \quad \tmu=-\frac{(\g_1 T_2 Z_2 + \g_2 T_1 Z_1) Z_3 V}{\tV^2}\,,
\end{equation}
\begin{equation}\label{eq:tildesimp2}
 \quad \tW_1=\frac{N_1}{T_1 T_2 V-\g_1 \g_2 Z_3}\,,\quad \tW_2=\frac{N_2}{T_1 T_2 V-\g_1 \g_2 Z_3}\,, \quad \tW_3=\frac{N_3}{T_1 T_2 V+\g_1 \g_2 Z_3}\,,
 \end{equation}
and we write
\begin{equation}\label{eq:tildesimp2}
 \tP_1=\frac{G_1}{N_1} +\frac{\tmu}{\tW_1}\,, \quad \tP_2=\frac{G_2}{N_2} +\frac{\tmu}{\tW_2}\,, \quad \tP_3=\frac{G_3}{N_3} +\frac{\tmu}{\tW_3}\,,
\end{equation}
where
\begin{equation}\label{eq:tildesimp3}
G_1=T_1 K_1 Z_1 V+\g_1 Z_2 Z_3\,, \quad G_2=T_2 K_2 Z_2 V+\g_2 Z_1 Z_3\,, \quad G_3=K_3 Z_3 V\,.
\end{equation}
We recall the form of the gauge potentials~\eqref{eq:AI}
\begin{align}
\begin{split}
 &\tA_I= \widetilde w_I+ \tP_I(d\psi + \tA)-\frac{dt+\tk}{\tW_I}\,.
\end{split}
\end{align}

To give the explicit form of $\widetilde F^{(3)}$ we trade all five-dimensional Hodge stars for three-dimensional Hodge stars and subsequently use the almost-BPS equations~\eqref{eq:almostBPS0}-\eqref{eq:almostBPS4} to replace as many of the three-dimensional Hodge stars as possible.

\paragraph{First term: $-\left(\frac{\widetilde {Z}^5_2}{\widetilde {Z}^2_3\widetilde {Z}^3_1}\right)^{1/4}\star_5 d\widetilde A_2$.}
We want to express
\begin{equation}
\star_5 d\widetilde A_2=\star_5 \left( d \widetilde w_2+ d\tP_2 \wedge (d \psi+\tA) +\tP_2  d\tA+ -d\left(\frac{1}{\tW_2}\right) \wedge (d t+\tk)-\frac{d\tk}{\tW_2}\right)\,,
\label{eq:starF2}
\end{equation}
whose five-dimensional Hodge stars are with respect to the metric
\begin{equation}
ds_5^2=-\frac{1}{\tZ_3 \sqrt{\tZ_1 \tZ_2}} (d t+\tk)^2+\sqrt{\tZ_1 \tZ_2} \Big(\tV^{-1}(d\psi+\tA)^2+\tV dy^i dy^i)\Big)\,,
\end{equation}
in term of three-dimensional Hodge stars with respect to the flat three-dimensional base spanned by $y_i$ with $i=1,2,3$.
We introduce the vielbeine
\begin{equation}
\te^0=f_0 (d t+\tk)\,, \qquad \te^1=f_1 (d\psi+\tA)\,, \qquad \te^i=f_i dy_i\,,
\end{equation}
where
\begin{equation}
f_0=(\tZ_1 \tZ_2)^{-1/4} \tZ_3^{-1/2}\,, \qquad f_1=(\tZ_1 \tZ_2)^{+1/4} \tV^{-1/2}\,,\qquad f_i=(\tZ_1 \tZ_2)^{+1/4} \tV^{+1/2}\,.
\end{equation}
With
\begin{eqnarray}
\star_5 (\te^i \wedge \te^j) &=& f_0 f_1 \te^0 \wedge \te^1 \wedge \epsilon_{ijk} \te^k\,,\nn\\
\star_5 (\te^i \wedge \te^0) &=& 
-f_1 f_i \te^1 \wedge \frac{\epsilon_{ijk}}{2} \te^j \wedge \te^k\,,\\
\star_5 (\te^1 \wedge \te^i) &=& f_0 f_i \epsilon_{ijk} \te^j \wedge \te^k \wedge \te^0\,,\nn
\end{eqnarray}
and thus
\begin{eqnarray}
\star_5 (dy^i \wedge dy^j) &=& f_0 f_1 f_i^{-1} (d t+\tk) \wedge (d \psi+\tA) \wedge \epsilon_{ijk} dy^k \,,\nn\\
\star_5 (dy^i \wedge (d t+\tk)) &=& -f_0^{-1} f_1 f_i (d \psi+\tA) \wedge \frac{\epsilon_{ijk}}{2} dy^j \wedge dy^k\,,\\
\star_5 ((d \psi+\tA) \wedge dy^i) &=& f_0 f_1^{-1} f_i \epsilon_{ijk} dy^j \wedge dy^k \wedge (d t+\tk)\,,\nn
\end{eqnarray}
the terms in~\eqref{eq:starF2} can be expressed as
\begin{eqnarray}
\star_5 d\widetilde w_2&=&f_0 f_1 f_i^{-1} (d t+\tk) \wedge (d \psi + \tA) \star_3 d\widetilde w_2\,,\nn\\
\star_5 \left[d\tP_2 \wedge (d\psi + \tA) \right]&=&-f_0 f_1^{-1} f_i (d t+\tk) \wedge \star_3 d\tP_2\,,\nn\\
\tP_2 \star_5 d\tA&=&  f_0 f_1 f_i^{-1} (d t+\tk) \wedge (  d \psi+\tA) \wedge \tP_2 \star_3 d\tA\,,\\
 \star_5 \left[-d\left(\frac{1}{\tW_2}\right) \wedge (d t+\tk) \right] &=& - f_0^{-1} f_1 f_i (  d \psi+\tA) \wedge \frac{\star_3 d\tW_2}{\tW_2^2}\,,\nn\\
- \star_5 \frac{d\tk}{\tW_2} &=& f_0 f_1^{-1} f_i (d t+d\tk) \wedge \frac{\star_3 d\tmu}{\tW_2} - f_0 f_1 f_i^{-1} (d t+d\tk) \wedge (d \psi+\tA) \frac{\tmu \star_3 d\tA}{\tW_2}\,.\nonumber
\end{eqnarray}
With this, the Hodge star combinations appearing in~\eqref{eq:starF2} are
\begin{equation}
\star_3 d\tA=-T_1T_2d V - \gamma_1\gamma_2 d Z_3\,,  \quad \star_3 d\widetilde w_2= K_2 T_1 dV + \gamma_1 dZ_3\,,
\end{equation}
\begin{equation}
\star_3 d\tW_2= \left(V\star_3d Z_3 - Z_3\star_3d V\right)\frac{\gamma_1\gamma_2T_1^2Z_2 + T_1T_2\gamma_2^2Z_1}{\widetilde{V}^2} +\frac{\gamma_2^2Z_3}{\widetilde{V}}\star_3d Z_1 +\frac{T_2^2V}{\widetilde{V}}\star_3d Z_2\,,
\end{equation}
and
\begin{align}
\star_3d \widetilde P_2-\frac{\star_3d\widetilde \mu}{\tW_2}&= \frac{N_1 \g_2 T_2}{N_2\widetilde{V}^2}\left(V\star_3 d Z_3-Z_3\star_3 d V \right) +\frac{VZ_3}{T_1T_2V-\g_1\g_2Z_3}\left(\g_1T_2\star_3 d Z_2+\g_2T_1\star_3 d Z_1\right)\,.
\end{align}

Finally we get for~\eqref{eq:starF2}:
\begin{align}\label{eq:ZZZF2}
-\Big(\frac{\tZ_2^5}{\tZ_1^3 \tZ_3^2}\Big)^{1/4} \star_5 d\widetilde A_2& = {(d t+\omega)\wedge(  d\psi+\tA)}\wedge \frac{\g_2 T_1 Z_1 Z_3  {d V}- \g_1 T_2 Z_2 V  {d Z_3}}{\g_1^2 V Z_2 Z_3^2 + T_1^2 Z_1 Z_3 V^2} \\
 &+ \frac{(d t+\tk)}{V Z_3 (\g_1^2 Z_2 Z_3 + T_1^2 Z_1 V)} \wedge \Big[\g_2 T_2  (\g_1^2 Z_2 Z_3+T_1^2 Z_1 V)(V  {\star_3 d Z_3}-Z_3  {\star_3 d V})\nn\\
 &\qquad\qquad\qquad \qquad \quad +V Z_3 \left(T_1 T_2 V - \g_1 \g_2 Z_3\right)(\g_1 T_2  {\star_3 d Z_2} + \g_2 T_1  {\star_3 d Z_1})\Big] \nonumber\\
 &+ \frac{( d \psi+\tA)}{\left(T_1 T_2 V - \g_1 \g_2 Z_3\right)^2} \wedge \Big[\g_2 T_2(\g_1 T_2 Z_2+\g_2 T_1 Z_1)(V  {\star_3 d Z_3}-Z_3  {\star_3 d V}) \nn\\
 &   \qquad\qquad \qquad\qquad \qquad  +\left(T_1 T_2 V - \g_1 \g_2 Z_3\right)(\g_2^2 Z_3  {\star_3 d Z_1}+T_2^2 V  {\star_3 d Z_2})\Big] \,.\nonumber
\end{align}

\paragraph{Second term: $d  \widetilde A_1 \wedge (  \widetilde A_3 -d  z )$.}
The second term in equation~\eqref{F3_short} is given by
\begin{eqnarray}\label{eq:A13z}
 d  \widetilde{A}_1 \wedge (  \widetilde{A}_3-d  z) &=&   d  \left[-\frac{1}{\tW_1}( d  t +\widetilde{k})+ \widetilde{P}_1(   d  \psi+\widetilde{A})+\widetilde w_1\right] \nn\\ 
 &&   \qquad \wedge  \left(-\frac{1}{\widetilde{W}_3}( d  t +\widetilde{k})+ \widetilde{P}_3(   d  \psi+\widetilde{A})+\widetilde w_3-d  z\right)\,.
\end{eqnarray}
Using~\eqref{eq:tildesimp1}~-~\eqref{eq:tildesimp3} we can write~\eqref{eq:A13z} as
\begin{align}\label{eq:A13zfinal}
d\tA_1 \wedge (\tA_3-d z)=
& -{(d t+\omega) \wedge (\widetilde w_3-d z)} \wedge d\left(\frac{1}{\tW_1}\right)+ {(  d \psi+\tA) \wedge (\widetilde w_3-d z)} \wedge  {d\Big(\frac{G_1}{N_1}\Big)}\nn\\
&+   {(d t+\omega) \wedge (  d \psi+\tA)} \wedge \Big[ \frac{1}{\tW_3}  {d\Big(\frac{G_1}{N_1}\Big)}-\frac{G_3}{N_3}\  d\Big(\frac{1}{\tW_1}\Big)\Big] \\
&+  \Big[ {(\widetilde w_3-d z)}-\frac{1}{\tW_3}  {(d t+\tk)} +\left(\frac{G_3}{N_3}+ \frac{\tmu}{\tW_3}\right)  {(  d\psi+\tA)}\Big] \wedge \Big[\frac{G_1}{N_1}  {d \tA}+ {d\widetilde w_1}\Big]\,,\nn
\end{align}
where, using the almost-BPS equations~\eqref{eq:almostBPS0}-\eqref{eq:almostBPS4} one can write
\begin{equation}
 d \widetilde A = -T_1T_2 \star_3d V-\gamma_1\gamma_2\star_3d Z_3\,,\quad 
 d \widetilde \omega_1= K_1T_2\star_3d V+\gamma_2 \star_3d Z_3\,.
\end{equation}
Putting~\eqref{eq:ZZZF2} and~\eqref{eq:A13zfinal} together we arrive at the following expression for the R-R field strength:
\begin{align}\label{eq:F3appendix}
&\widetilde F^{(3)} = 
 \left[(dt+\omega)\wedge d\left(\frac{T_1 T_2 V - \g_1 \g_2 Z_3}{N_1}\right) - (d\psi+\tA) \wedge d\Big(\frac{T_1 K_1 Z_1 V+\g_1 Z_2 Z_3}{N_1}\Big) \right]  \wedge (\widetilde w_3-dz)  \nonumber\\
& +  {(dt+\omega) \wedge (d\psi+\tA)} \wedge \left[\frac{T_1 T_2 V+\g_1 \g_2 Z_3}{N_3}{d\left(\frac{T_1 K_1 Z_1 V+\g_1 Z_2 Z_3}{N_1}\right)} - K_3 d\left(\frac{|T_1 T_2 V - \g_1 \g_2 Z_3|}{N_1}\right)\right] \nonumber\\
& + \left[ {(\widetilde w_3-dz)}-\frac{T_1 T_2 V+\g_1 \g_2 Z_3}{N_3} (dt+\omega) +K_3  {(d\psi+\tA)}\right] \wedge \left[\frac{T_1 K_1 Z_1 V+\g_1 Z_2 Z_3}{N_1}  {d\tA}+ {d\widetilde w_1}\right]\nonumber\\
& + {(dt+\tk)}\wedge \Big[ \frac{\g_2 T_2}{N_3}  (V  {\star_3 dZ_3}-Z_3  {\star_3 dV}) +\frac{T_1 T_2 V - \g_1 \g_2 Z_3}{N_1}(\g_1 T_2  {\star_3 dZ_2} + \g_2 T_1  {\star_3 dZ_1})\Big]  \nonumber\\
& +(d\psi+\tA)\wedge \Big[\frac{\g_1 \g_2 T_2^2 Z_2+\g_2^2 T_1 T_2 Z_1}{(T_1 T_2 V - \g_1 \g_2 Z_3)^2}(V  {\star_3 dZ_3}-Z_3  {\star_3 dV}) +\frac{\g_2^2 Z_3  {\star_3 dZ_1}+T_2^2 V  {\star_3 dZ_2}}{T_1 T_2 V - \g_1 \g_2 Z_3}\Big]\nonumber\\
& +(dt+\omega)\wedge(d\psi+\tA) \wedge \frac{\g_2 T_1 Z_1 Z_3  {dV}- \g_1 T_2 Z_2 V  {dZ_3}}{N_1 N_3}\,,
\end{align}
where 
\begin{align}
\begin{split}
& \qquad\quad \qquad \widetilde w_1=w_1+\g_2 v_3\,, \quad \widetilde w_3=w_3+\g_1 v_2 +\g_2 v_1 - \g_1\g_2\nu\,, \\
&\tk=-\frac{V Z_3(\g_1T_2Z_2+\g_2 T_1Z_1)}{\tV^2}(d\psi + \tA)\,,\quad \tA=A-\g_1 w_1 - \g_2 w_2 - \g_1\g_2 v_3\,,
\end{split}
 \end{align}
 and we recall from~\S~\ref{subsec:completion}:
 \begin{eqnarray}
  N_1=\g_1^2 Z_2 Z_3 + VT_1^2 Z_1\,,\quad N_2=\g_2^2 Z_1 Z_3+VT_2^2Z_2\,,\quad N_3 = VZ_3\,.
 \end{eqnarray}
As a check, a tedious but straightforward exercise shows that the expression~\eqref{eq:F3appendix} is closed.

}

\bibliographystyle{toine}
\bibliography{References.bib}

\end{document}